\documentstyle[12pt]{article}
\begin{document}
\centerline{\bf ON  THE  GLASS  TRANSITION  TEMPERATURE}
\centerline{}
\centerline{\bf IN  COVALENT  GLASSES}
\centerline{}
\centerline{by} 
\vskip 0.5cm
\centerline{Richard Kerner  and  Matthieu Micoulaut}
\vskip 1cm
\centerline{\it Laboratoire GCR - UFR de Physique, Paris VI, CNRS - URA 769}
\centerline{\it Universit\'e Pierre et Marie Curie, Tour 22, Boite 142}
\centerline{\it 4, Place Jussieu, 75005 Paris, France}
\vskip 0.6cm
\indent
{\bf Abstract:}
\newline
\indent
{\em We give a simple demonstration of the formula relating the glass 
transition temperature, $T_g$, to the molar concentration $x$ of a modifier 
in two types of glasses: binary glasses, whose composition 
can be denoted by $X_n Y_m + x\, M_p Y_q$, with X an element of $III$-rd or 
$IV$-th group (e.g. $B$, or $Si$, $Ge$), while $M_p Y_q$ is an alkali 
oxide or chalcogenide; next, the network glasses of the type 
$A_x\,B_{1-x}$, e.g. $Ge_x\,Se_{1-x}, \, Si_x\,Te_{1-x}$, etc. After
comparison, this formula gives an exact expression of the parameter $\beta$ 
of the modified Gibbs-Di Marzio equation.}
\newpage
\section{Introduction}
The nature of glass transition is complex \cite{r1} and even today remains 
poorly understood. Numerous studies \cite{r3} have been devoted to 
measurements and understanding of the glass transition temperature, $T_g$, 
which is influenced by experimental conditions (e.g. the 
cooling rate of the melt). Nevertheless, even for the measurements under 
standard conditions, there is still no consensus as to which structural or 
thermodynamical factors are responsible for determining $T_g$. Kauzmann has 
observed that many glass-formers have $T_g\sim {\frac {2}{3}}T_{melting}$ 
\cite{r3b}. Other attempts suggested correlation with molecular or atomic 
motions with characteristic energies that might be responsible for the 
observed values of $T_g$ \cite{r3c}. More recently, Tanaka discovered an 
empirical relationship between $T_g$ and the average coordination number $m$ 
per atom in glass : $ln\ T_g\sim 1.6 m + 2.3$ \cite{r3d}.
\par
Nevertheless, all these descriptions become more and more complicated if one 
starts to vary the composition $x$ of glass forming material. Typical examples 
are the chalcogenide glasses (e.g. $As_xSe_{1-x}$) or binary glasses involving 
a network former (e.g. silica) and a modifier (an alkali oxide or chalcogenide, 
e.g. $Li_2O$). Some simple rules for predicting the glass transition 
temperature under standard conditions as function of the composition $x$ and 
the nature of the atoms involved exist \cite{r4} among which Tanaka's 
relation \cite{r3}, but there is still no general formulation of $T_g$
versus $x$ which could give the precise shape and explain with a 
mathematical model
why maxima in glass transition temperature can occur (as in $B_2O_3$ based 
systems \cite{r5} or in $Ge_xSe_{1-x}$ glasses at x=0.33 \cite{r6}).
\par
For many structural glasses, like the alkali-borate $(1-x)\,B_2O_3 
+ x\,Li_2O$ glass, or the selenium-germanium network glass $Ge_x Se_{1-x}$,
at least at low concentrations of the modifier, their glass transition 
temperature increases as even small amounts of modifier are added to the melt.
In many cases the function, $T_g(x)$, is quasi-linear only for low modifier 
concentrations (small values of $x$), and in some cases (e.g. the 
aforementioned $B_2O_3$ and $Ge_xSe_{1-x}$ based glasses) displays a peak at 
some value of $x$, after which the derivative, $d\,T_g / d\,x$, changes its 
sign and becomes negative. An elegant explanation of this phenomenon, based on
the evaluation of average number of constraints per structural unit, has been
proposed for the $Ge_x Se_{1-x}$ alloy by J.C. Phillips (\cite{r7}).
\par
We propose in this paper a simple and general model of glass formation
enabling us to derive the dependence of $T_g$ on $x$, and a very simple 
formula for its derivative at $x = 0$. In the refs.\cite{r8} and \cite{r9} we 
have derived this formula
and applied it to the particular case of alkali-borate glass; in this paper
we show that it is confirmed by a large set of experimental data, concerning
mainly network glasses.
\par
It seems obvious that a liquid which is about to undergo the glass 
transition, during which its viscosity grows exponentially, is composed of 
many clusters, various in size and shape, agglomerating as the 
temperature decreases. Usually the process of solidification leading to the 
vitreous state, interpreted as a {\it continuous random network}, (the term 
introduced by Zachariasen, \cite{r10}) occurs in a small temperature range, 
(often a function of the cooling rate), between $30$ and $40 K$ around the 
average value called the {\it glass transition temperature} $T_g$ \cite{r11}.
It is known that the process is not a regular second order phase transition 
in the thermodynamical sense \cite{r12}.
\par
The most important elementary process of agglomeration, which is the creation
of a new bond (usually an oxygen, sulfur or selenium bridge) between two 
clusters, can be analyzed in the ease of binary glasses. The
modifier transforms the atoms of the glass former in a well defined manner 
(which may change when the concentration of the modifier increases); e.g., when a 
small amount of $Li_2O$ is added to $B_2O_3$ glass, the ions of lithium 
transform the threefold boron atoms of the network former into fourfold 
boron atoms \cite{r5}; when the concentration $x$ attains about $35\%$, the 
action of the modifier changes, and new lithium ions begin to depolymerize
the network (by creating non-bridging oxygens), transforming the threefold 
boron atoms so that they have only $2$ free bonds left that can 
participate in network formation. A similar modification of the atoms of 
glass former occurs in many other glasses; almost always, it amounts to the 
change of the coordination number of the elementary basic structure unit of 
the glass former (e.g. four-, three-, two-, one- and zero-fold tetrahedra in 
$IV-VI$ based glasses as $SiS_2-Li_2S$ systems, corresponding to
$SiS_4^{n\ominus}$ [n=4..0] units \cite{r13}).
\section{The model}
When the modifier's concentration is not too high, we assume that the 
probability of finding an "altered" atom with coordination number
$m'$ (such as a germanium atom in
$Ge_xSe_{1-x}$ systems) on the rim of an average cluster is 
equal to $x$, and probability of finding a "regular" atom 
with coordination number $m$ is $1-x$
(such as a selenium atom in $Ge_xSe_{1-x}$). Let us 
denote the non-altered ("regular") atoms of the glass former by $A$, and the 
modified ("altered") atoms by $B$.
\par
Our aim is to evaluate the time dependence of the fluctuations of local concentration 
(i.e. local variations of the variable $x$), and derive the equations imposing
the minimization of these fluctuations. This minimum should then correspond 
to a stable or meta-stable configuration \cite{r14,r15}. With only two types 
of atoms there 
are three elementary processes of single bond formation, A-A, A-B and B-B.
\par
The probabilities of these processes are proportional to the products of the
relative concentrations of the atoms involved, the statistical factors which
may be regarded as the degeneracies of the corresponding energies, and are 
proportional to the products of the corresponding valencies, and the Boltzmann 
factors involving the energies of respective bond formation. If we look 
at the newly created pairs, their probabilities will be:
\begin{equation}
\label{1}
p_{AA1} = \frac{m^2}{Q}\,(1 - x)^2\,e^{-\frac{E_1}{kT}}
\end{equation}
\begin{equation}
\label{2}
p_{AB1} =\frac{2mm'}{Q}\,x(1 - x)\,e^{-\frac{E_2}{kT}}
\end{equation}
\begin{equation}
\label{3}
p_{BB1}= \frac{{m'}^2}{Q}\,x^2\,e^{-\frac{E_3}{kT}}
\end{equation}
where $Q$ is the normalizing factor given by
\begin{equation}
\label{4}
Q = m^2\,(1-x)^2\,e^{-\frac{E_1}{kT}} + 2m m'\,x\,(1-x)\,e^{-\frac{E_2}{kT}} 
+ {m'}^2\, x^2\,e^{-\frac{E_3}{kT}}
\end{equation}
\indent
For the moment, we exclude the possibility of a simultaneous creation of two
bonds leading to the formation of four- and six-membered rings; such 
posibility does exist in several glasses, and can be taken into account; 
then the number of elementary processes is bigger, but the analysis remains 
exactly the same \cite{r10,r11}. Anyhow, the
network glasses we are considering here do not possess such an ability.
Although the overall averages remain the same, the new pairs may create a
local fluctuation in the statistics, which can be evaluated as:
\begin{equation}
\label{5}
x^{(1)} = \frac{1}{2}\,( 2\,p_{BB1} + p_{AB1} )
\end{equation}
\indent
Denoting the average time needed to form a new bond by  $\tau$, we can 
approximate the time derivative of $x$ due to the above fluctuation as
\begin{equation}
\label{6}
\frac{d\,x}{d\,t} \simeq \frac{1}{\tau}\,[\, x^{(1)} - x\,]
\end{equation}
which may be considered a good approximation as long as the time variation
of the temperature, $dT/dt$, can be neglected, which is true when the quenching
rate, $q = dT/dt$, satisfies:
\begin{equation}
\label{7}
\frac{1}{T}\,\frac{d\,T}{d\,t} \, = \, \frac{1}{T}\,q \ll \frac{1}{\tau}
\end{equation}
\indent
This inequality is particularly well satisfied for the systems with which
we shall be dealing.
For example, borate and silicate based glasses form very easily, and have
critical cooling rates of the order of $q_{crit} \simeq 10^{-4} K\,.s^{-1}$
\cite{r16}; so do other network glasses we are going to investigate \cite{r17}.
\newline
\indent
Now, if we want to {\it minimize} the local fluctuations, we should equate the
above expression to $0$, which amounts to finding the {\it stationary} or
{\it singular} solutions of the differential equation (\ref{6}) This leads to:
\begin{equation}
\label{8}
\frac{d\,x}{d\,t} \simeq {\frac {1}{\tau}}\biggl(p_{BB1} + {\frac {1}{2}}
p_{AB1}-x \biggr)= 0
\end{equation}
which leads to a simpler condition:
\begin{equation}
\label{9}
x (1 - x) \biggl[ \,m\,(1-x)\,(m'\,e^{-\frac{E_2}{kT}}- m\,e^{-\frac{E_1}{kT}}) 
+ m'\,x\,(m'\,e^{-\frac{E_3}{kT}} - m\,e^{-\frac{E_2}{kT}})\,\biggr] = 0,
\end{equation}
There are always two singular solutions at the points $x = 0$ and $x = 1;$ 
there can exist also a third solution, given by the following expression:
\begin{equation}
\label{10}
x_{am} = \frac{m m'\,e^{-\frac{E_2}{kT}} - m^2\, e^{-\frac{E_1}{kT}}}
{2\,mm'\,e^{-\frac{E_2}{kT}} - m'^2\, e^{-\frac{E_3}{kT}} - m^2\, 
e^{-\frac{E_1}{kT}}}
\end{equation}
However, recalling that the variable $x$ represents the modifier concentration,
it is physically acceptable only if $0 \leq x_{am} \leq 1;$ moreover, we want
it to be an {\it attractive point,} which is possible only if:
\begin{equation}
\label{11}
\frac{m'}{m} > e^{\frac{E_2 - E_1}{kT}} \, \ \ {\rm and} \ \
\frac{m}{m'} > e^{\frac{E_2 - E_3}{kT}} , 
\end{equation}
\indent
We interpret the presence of the stable solution, $x_{am}$, as the 
manifestation of tendency of the system to become amorphous.
On the contrary, when condition (\ref{11}) is not satisfied, the stable
({\it attractive}) solution is found at $x = 0$, which means that at the
microscopic level the agglomeration tends to separate the two kinds of atoms,
$A$ and $B$.
\par
Note that due to the homogeneity of the expression (\ref{10}), only {\it two}
energy differences are essential here: $\alpha_1 = (E_2 - E_1)/kT$, and 
$\alpha_2 =(E_2 - E_3)/kT.$
\par
Various glass formers display the tendency towards the solution $x_{am}$ in
a wide range of modifier concentration, also when it tends to zero 
($x \rightarrow 0$). In this
limit we get one condition only, which concerns only the energy 
difference $E_2 - E_1:$
\begin{equation}
\label{12}
E_2 - E_1 = k\,T_0\,ln(\frac{m'}{m}) .
\end{equation}
where $T_0$ is the glass transition temperature at $x = 0$, i.e. in the limit
when the modifier concentration goes to zero.
\newline
\indent
This equation summarizes the relation between the {\it entropic} and 
{\it energetic} factors that are crucial for the glass forming tendency to
appear. It tells us that in good binary glass formers whenever $m' >m$ 
(and $ln(\frac{m'}{m}) > 0$), one must have $E_2 > E_1$, and vice versa.
\par
This condition is what should be intuitively expected indeed when a system displays the 
tendency towards amorphisation: it behaves in a "frustrated" way in the sense 
that the two main contributions to the probabilities of forming bigger 
clusters act in the opposite directions. Whenever the modifier raises the
coordination number ($m' > m$), thus creating more degeneracy of the 
given energies $E_i$ (i.e. much more possiblities of linking
two entities A and B)
and increasing the probability of agglomeration, the corresponding Boltzmann 
factor $e^{-\frac{E_2}{kT}}$ is smaller than for the non-modified atoms,
$e^{-\frac{E_1}{kT}}$, reducing the probability of agglomeration, and
vice versa.
\par
Recalling that the stable solution corresponding to the glass-forming tendency
defines an implicit function, $T(x)$, via the relation
\begin{equation}
\label{13}
 \,m\,(1-x)\,(m'\,e^{-\frac{E_2}{kT}}- m\,e^{-\frac{E_1}{kT}}) + m'\,x\,
(m'\,e^{-\frac{E_3}{kT}} - m\,e^{-\frac{E_2}{kT}})\, = \, \Phi(x, T) = 0,
\end{equation}
\indent
we can easily evaluate the derivative of $T$ with respect to $x$:
\begin{equation}
\label{14}
\frac{d T}{d x}\, = -\,\biggl[\biggl( \frac{\partial \Phi}{\partial x}\biggr)/
\biggl(\frac{\partial \Phi}{\partial T}\biggr) \biggr]_{\Phi(x,T)=0} 
\end{equation}
\indent
with $m$ and $m'$ representing the average number of free valencies of the
two species of involved atoms.
\par
In the limit $x = 0$ the result has a particularly simple form:
\begin{equation}
\label{15}
\biggl[ \frac{d T}{d x} \biggr]_{x=0} \,= \, \biggl[\frac{d\,x_{am}}{d\,T}
\biggl]^{-1}_{T=T_0} \, = \biggl[ \frac{(2\,(m'/m)\,e^{\frac{E_1-E_2}{kT}} - 1)^2} 
{\frac{E_2-E_1}{kT^2}} \biggr]_{T=T_0} 
\end{equation}
Inserting condition (\ref{12}), we obtain a general relation which can be 
interpreted as a universal law:
\begin{equation}
\label{16}
\biggl[\frac{dT}{dx} \biggr]_{x=0} \, = \, \frac{T_0}{ln(\frac{m'}{m})} 
\end{equation}
The glass transition temperature $T_g$ in binary glasses increases
with the addition of a modifier that increases the average coordination number
($m' > m$), and decreases with the addition of a modifier that decreases the
average coordination number ($m' < m$).
\section{Discussion}
The formula (\ref{16}) can be quite easily compared with experimental data. In Reference \cite{r9}, 
we have demonstrated a good agreement of this formula with the behavior
of $dT_g/dx$ at $x=0$ for the alkali-borate glass $(1-x)\,B_2O_3 \,x\,Li_2O$
and the silicate glass $(1-x)\,SiO_2\,+\,x\,CaO$. Formula (13) predicted
$m'/m \simeq 3/2$ in the first case, which leads to a positive derivative, 
and $m'/m \simeq 1/3$ giving a negative one in the second case.
\par
We have tried to find, whenever possible, reported glass transition temperatures
of glass-forming systems which were composed of a very high fraction of 
"regular" atoms (e.g. selenium, sulphur or tellurium in the case
of chalcogenide glasses), i.e. close to $95\%$, in order to approach 
the limit value $x\rightarrow 0$ of formula
(\ref{16}). We have applied this model to simple glass forming 
systems, namely chalcogenide based glasses, for which numerous experimental 
data are available. 
Within the ranges (close to $x=0$) found in various references (see Table I 
and II), our formula agrees very well with the experimental data for the 
selenium based glasses (Table I). By "agreement", we mean the fact that
the computation of the slope of $T_g$ at $x=0$ obtained with the experimental 
measured values, leads to very satisfying values of $m'/m$. 
Starting from pure vitreous 
selenium whose average glass transition temperature is $316\, K$ (obtained by
extrapolation from different experimental measurements \cite{r12}),
we observe that all the considered glasses display an increase of the glass
transition temperature with increasing $x$ . Indeed, the {\it two valenced}
selenium atoms form a network of chains with various length \cite{r18}, 
hence $m = 2$ in this case. The atoms of the modifier ($\ Ge,\ Si,\ As\ $)
produce cross-linking between the $Se$-chains, thus
creating new stable structural units whose $m'$ is equal to $4$ ($Si,\,Ge$) 
or $3$ ($As$) \cite{r19}.
Obviously, equation (\ref{16}) agrees only for the chalcogen-rich region, 
in particular for germanium selenide glasses.
A similar phenomenon is observed in the $Te$- and $S$-based glasses. The average
value of $T_g$ of vitreous tellurium, obtained by averaging and extrapolation,
is about $343\,K$ \cite{r3d}, whereas the $T_g$ of pure vitreous sulphur is 
about $245\,K$ \cite{r12}.
\par
The tables below show a satisfactory agreement with our formula, and prove
beyond any doubt that it is not a matter of coincidence.\par
\vspace{0.5cm}
\begin{tabular}{|c||c|c||c|c|c|c|}\hline
\centering
Compound&$\biggl( {\frac {m'}{m}}\biggr)_{th}$&$\biggl({\frac {m'}{m}}\biggr)_{exp}$
&\multicolumn{2}{c|}{Obtained with}&$\Delta T_g [K]$&Reference \\
 & & &$x$&$T_g [K]$& & \\ \hline\hline
$Ge_xSe_{1-x}$&2.0&2.04&0.05&336&22&\cite{r6} \\ \hline
$Si_xSe_{1-x}$&2.0&2.04&0.05&336&22&\cite{r20} \\ \hline
$As_xSe_{1-x}$&1.5&1.54&0.003&318&2&\cite{r21} \\ \hline
$Sb_xSe_{1-x}$&1.5&1.31&0.15&493&177&\cite{r22} \\ \hline
$P_xSe_{1-x}$&2.5&2.53&0.05&333&17&\cite{r23} \\ \hline
\end{tabular}
\par
\vspace{0.5cm}
{\em Table I: Different selenium based glasses. Comparison between the 
theoretical value of $m'/m$ and the experimental value deduced from the slope 
using data of $T_g$ for the lowest available concentration $x$.}\par
\vspace{0.5cm}
\begin{tabular}{|c||c|c||c|c|c|c|}\hline
\centering
Compound&$\biggl({\frac {m'}{m}}\biggr)_{th}$&$\biggl({\frac {m'}{m}}\biggr)_{exp}$&\multicolumn{2}{c|}{Obtained with}&$\Delta T_g [K]$&Reference \\
 & & &$x$&$T_g [K]$& & \\ \hline \hline
$Si_xTe_{1-x}$&2.0&2.11&0.10&389&46&\cite{r24} \\ \hline
$Ge_xTe_{1-x}$&2.0&1.97&0.15&419&76&\cite{r25} \\ \hline
$Ga_xTe_{1-x}$&1.5&1.45&0.20&528&185&\cite{r26} \\ \hline \hline
$As_xS_{1-x}$&1.5&1.54&0.11&307&63&\cite{r27} \\ \hline
$Ge_xS_{1-x}$&2.0&1.72&0.10&290&45&\cite{r28} \\ \hline
\end{tabular}
\par
\vspace{0.5cm}
{\em Table II: Different tellurium and sulphur based glasses. Comparison 
between the theoretical value of $m'/m$ and the experimental value of $m'/m$ 
deduced from the slope using data of $T_g$ for the lowest available 
concentration $x$.}\par
\vspace{0.5cm}
Nevertheless, we should mention that in certain glasses the equation (\ref{16}) 
does not seem to be well satified. We believe that this problem is mostly due to the fact that
certain modifiers exhibit already a metallic character; this may explain why
our formula can not be applied to e.g. $Al_x Te_{1-x}$ glasses \cite{r29}. 
The covalent bonding with well-defined valencies is not well suited for the 
description of the network in this case. 
\newline
\indent
A different type of problem arises in the systems which do not form glass
below certain minimal value of $x$; it is obvious that our formula does not
apply when glass can not be formed in the limit $x = 0$. Nevertheless, the
formula can sometimes be extrapolated down to $x=0$, when the variation
of $T_g$ versus $x$ represents a straight line for greater values of $x$.
The constant slope allows then a comparison with (\ref{16}).
In order to test the universality of the relation (\ref{16}), experimental 
measurements
in the very low $x$ concentration range remain still to be realized (as 
for $As_xSe_{1-x}$, Table I) and also on the compounds which to our 
knowledge have never
been investigated, such as $B_{.05}Se_{.95}$ and for which we predict 
$T_g\sim 355K$, and $B_{.05}S_{.95}$ for which we predict $T_g\sim 275K$.

\section{Relationship with the Gibbs-Di Marzio equation}

At the beginning of this article, we have mentioned that empirical 
relationships
between glass transition temperature and several structural or physical
properties have been proposed during the past ten years \cite{r4}, \cite{r3d}.
Among these, we would like to focus our attention on the relation of 
equation (\ref{16}) with the so-called
"{\em Gibbs-Di Marzio}" equation which is particularly well adapted for
predicting $T_g$ in chalcogenide glasses. As described above (see figure 3), 
one can consider the chalcogenide glass system as a network of chains 
(e.g. selenium atoms) in which cross-linking units (such as germanium atoms)
are inserted. Gibbs and Di Marzio \cite{r30} have developed a theory,
based on equilibrium principles and assuming that glass transition is a
second-order phase transition, which relates the increase of $T_g$ to
the growing presence of these cross-linking agents. They have
applied successfully their theory to explain the $T_g$ data in polymers
\cite{r31,r32}. An adapted theory constructed by Di Marzio \cite{r33} has 
shown that for glass systems with some chain stiffness, the glass transition
temperature versus cross-linking density $X$ could be expressed as:
\begin{equation}
\label{17} 
T_g\ =\ {\frac {T_0}{1-\kappa X}}
\end{equation}
where $T_0$ is the glass transition temperature of the initial 
polymeric chain and $\kappa$ a constant.
\par
Later on, Sreeram and co-workers \cite{r4} have modified this equation and 
have expressed $T_g$ in terms of the network average coordination number
$<r>$ which is widely used for the description of network glasses
since Phillips has introduced the concept of $<r>$ in his constraint theory 
\cite{r34}.
These authors have redefined for multicomponent chalcogenide glasses the
cross-linking density $X$ as being equal to the average coordination
number of the cross-linked chain less the coordination number of the initial
chain, i.e.:
\begin{equation}
\label{18}
X\ =\ <r>\ -\ 2
\end{equation}
The Gibbs-Di Marzio equation may then be rewritten as:
\begin{equation}
\label{19}
T_g={\frac {T_0}{1-\beta (<r>-2)}}
\end{equation}
where $\beta$ is a system depending constant, related eventually to the
bond interchange \cite{r35} (which is responsible for the system dependent
structural relaxation), whereas it was suggested that the constant $\kappa$
is universal \cite{r33}. Sreeram et al. fitted (least-squares fit) the
constant $\beta$ to their $T_g$ measurements \cite{r4} and obtained the
value between:
\begin{equation}
\label{20}
0.55\ <\ \beta\ <\ 0.75
\end{equation}
depending on the considered system and the involved atoms.
\par
Let us now try to relate the modified Gibbs-Di Marzio equation to formula
(\ref{16}) in the pure chalcogen limit $x=0$, in order to give an analytical
expression of $\beta$. According to Phillips \cite{r34}, we have expressed
the average coordination number $<r>$ in terms of the coordination
number of the covalently bonded atoms given by the $8-N$ rule (where N is
the number of the outer shell electrons), i.e. our previously defined
factors $m$ and $m'$:
\begin{equation}
\label{21}
<r>\ =\ m(1-x)+m'x
\end{equation}
The slope at the origin, where $x=0$ (and $<r>=m$) is then:
\begin{equation}
\label{22}
\biggl[{\frac{d\ T_g}{d<r>}}\biggr]_{<r>=m}\ =\ {\frac {T_0}{(m'-m)\ ln{\frac {m'}{m}}}}
\end{equation}
which corresponds to a constant slope and to the straight line with equation:
\begin{equation}
\label{23}
T_g\ =\ T_0\biggl[1+ {\frac {1}{(m'-m)\ ln{\frac {m'}{m}}}}(<r>-m)\biggr]
\end{equation}
In the vicinity of the pure chalcogen region, the Gibbs-Di Marzio can be 
expressed as:
\begin{equation}
\label{24}
T_g\ \simeq\ T_0\biggl[1+\beta\ (<r>-2)\biggr]
\end{equation}
which leads by identifying the two latter equations to an analytical 
expression of the constant $\beta$, involving only the coordination number
$m'$ and $m$. In Ref. \cite{r4}, Sreeram started their investigation of
multicomponent chalcogenide systems from pure vitreous selenium, hence
$m=2$:
\begin{equation}
\label{25}
{\frac {1}{\beta}}\ =\ (m'-m)\ ln({\frac {m'}{m}})
\end{equation}
\par
\vspace{0.5cm}
\begin{tabular}{|c|c|c|c|} \hline
\centering
System&$\beta_{fit}$&Correlation&Reference \\ 
 & &coefficient& \\ \hline
$Ge_xSe_{1-x}$&0.74&0.993&\cite{r6} \\
$Ge_xSe_{1-x}$&0.72&0.988&\cite{r4} \\
$Ge_xSe_{1-x}$&0.65&0.993&\cite{r36} \\
$Ge_xS_{1-x}$&0.73&0.998&\cite{r28}\\
$Si_xSe_{1-x}$&0.81&0.997&\cite{r20}\\ \hline
\end{tabular}
\par
\vspace{0.5cm}
{\em Table III: Computed values of $\beta$ obtained from a least-squares 
fit, for different binary glass systems, compared to the value $\beta =0.72$
calculated from equation (\ref{25}).}
\vspace{0.5cm}

The value of $\beta$ can now be computed for binary systems if the coordination 
number of the atoms are known, e.g. for chalcogenide based glasses, the possible
values for $\beta$ are $0.36$ (for $m'=5$), $2.47$ (for $m'=3$) and $0.72$
(for $m'=4$). The latter situation corresponds to the glass $Ge_xSe_{1-x}$
and the agreement of $\beta=1/2ln(2)=0.72$ with the value obtained by a 
least-squares fit of the glass transition temperatures data versus $<r>$, is 
very good. Other IV-II systems behave very similarly, as seen in Table III.
\section{Summary and conclusion}
We have derived in this paper an analytical formula from the statistical 
consideration of the agglomeration of coordinated entities (which may be
identified with atoms or valenced clusters), which gives the slope of the
glass transition temperature in very poorly modified binary systems 
$X_xY_{1-x}$ with $x\simeq 0$:
\begin{equation}
\label{26}
\biggl[\frac{dT}{dx} \biggr]_{x=0} \, = \, \frac{T_0}{ln(\frac{m'}{m})} 
\end{equation}
The comparison of this formula with previously computed relationships of
$T_g$ with the concentration $x$ or the average coordination number $<r>$, 
leads to an exact value of the $\beta$ parameter of the modified Gibbs-Di
Marzio equation of glass transition temperature.
\begin{equation}
\label{27}
{\frac {1}{\beta}}\ =\ (m'-m)\ ln({\frac {m'}{m}})
\end{equation}
The formula seems to agree with systems investigated elsewhere.
\par
We believe that such an attempt should be generalized for every concentration
in binary glasses, in order to explain by means of the involved coordination
numbers of the characterisitic clusters, the extrema of $T_g$ in different
systems such as $Ge_xSe_{1-x}$ or $B_2O_3$ based glasses.
Other direction of investigation should be the derivation of a similar 
formula for pseudo-binary or ternary glass systems, for which there is even
more experimental data available. This work will be the subject of
forthcoming papers.

\end{document}